\documentclass[aps, pra, twocolumn, noeprint, longbibliography, 10pt]{revtex4-2}
\usepackage{amsmath, bm, amssymb, dsfont, wasysym}  
\usepackage{indentfirst}  
\usepackage{graphicx}  
\usepackage{color}  
\usepackage{hyperref}  

\def\ie{\textit{i.e.}}

\def\ket#1{\left|#1\right\rangle}  
\def\bra#1{\left\langle #1\right|}  
\def\identity{\hat{\mathds{1}}}  
\def\Sop#1{\hat{j}_{#1}}  
\def\rhoopt{\hat{\rho}^{\prime}}  
\def\rhoin{\hat{\rho}_0}  
\def\rhoy{\hat{\rho}^y}  
\def\piopt{\pi^{\prime}}  
\def\Tpulse{T_{\mathrm{pulse}}}  
\def\OmegaRef{\Omega_{\mathrm{R}}}  

\hypersetup{
    colorlinks=true,
    linkcolor=blue,
    citecolor=blue,
    filecolor=blue,
    urlcolor=blue,
}

\begin{document}

\title{Reinforcement Learning for Optimal Control of Spin Magnetometers}

\author{Logan W. Cooke}
\author{Stefanie Czischek}
\affiliation{University of Ottawa, Department of Physics, 75 Laurier Ave E, Ottawa, ON, Canada}

\date{\today}

\begin{abstract}
    Quantum optimal control in the presence of decoherence is difficult, particularly when not all Hamiltonian parameters are known precisely, as in quantum sensing applications. 
    In this context, maximizing the sensitivity of the system is the objective, for which the optimal target state or unitary transformations are unknown, especially in the case of multi-parameter estimation.
    Here, we investigate the use of reinforcement learning (RL), specifically the soft actor-critic (SAC) algorithm, for problems in quantum optimal control.
    We adopt a spin-based magnetometer as a benchmarking system for the efficacy of the SAC algorithm.
    In such systems, the magnitude of a background magnetic field acting on a spin can be determined via projective measurements.
    The precision of the determined magnitude can be optimized by applying pulses of transverse fields with different strengths.
    We train an RL agent on numerical simulations of the spin system to determine a transverse-field pulse sequence that optimizes the precision and compare it to existing sensing strategies.
    We evaluate the agent's performance against various Hamiltonian parameters, including values not seen in training, to investigate the agent's ability to generalize to different situations.
    We find that the RL agents are sensitive to certain parameters of the system, such as the pulse duration and the purity of the initial state, but overall are able to generalize well, supporting the use of RL in quantum optimal control settings.
\end{abstract}

\maketitle


\section{Introduction}
\label{s:introduction}

As quantum systems find more and more applications in fields like quantum computation~\cite{divincenzo_1995, steane_1998, gyongyosi_2019}, quantum simulation~\cite{daley_2022, altman_2021}, quantum sensing~\cite{degen_2017}, or quantum imaging~\cite{defienne_2024}, their strengths beyond limitations of classical systems become increasingly imminent. 
However, to exhaust their full potential, quantum systems need to be controlled very precisely. 
This optimal control of quantum systems remains a difficult problem, particularly in the presence of decoherence effects and experimental imperfections, that need to be taken into account in the control process. 
The quality of control can be measured by the infidelity of target unitary transformations, which needs to be minimized to achieve a well-controlled implementation.
Furthermore, minimizing the duration of target unitary transformations avoids decoherence effects.
In both cases, optimal target states or unitaries must be known in advance.
Algorithms such as gradient ascent pulse engineering (GRAPE)~\cite{khaneja_2005}, or chopped random basis (CRAB)~\cite{caneva_2011, doria_2011} have been widely successful in optimizing for these objective functions. 
There are, however, more complicated objective functions for which to optimize, or situations in which only partial information about system parameters is known. 
In quantum sensing applications, for instance, the goal is to perform a multi-parameter estimation task with maximal sensitivity with respect to each unknown parameter, in the presence of decoherence effects~\cite{degen_2017, goldberg_2021, ansel_2024}. 
It is therefore of interest to find robust and inexpensive means of adaptive control.

Recent work has suggested algorithms from reinforcement learning (RL) as a potential solution to the problem of quantum optimal control~\cite{metz_2023, milson_2023, li_2025}.
These algorithms are designed to determine the optimal behavior in the face of uncertainty (typically applied in game theory) with excellent convergence and minimal resources. 
Examples in the field of quantum sensing include the optimization of a quantum kicked-top sensor~\cite{schuff_2020}, adaptive control feedback in a multi-parameter estimation task~\cite{cimini_2023}, and the optimization of an ultracold atom gyroscope~\cite{chih_2024}.
Here, we apply RL to optimize the control of a semi-classical sensor. 
The problem consists of training an agent to make decisions based on observations from the system. 
In this case, the agent will determine the particular control parameters to apply to a quantum state in order to maximize its sensitivity to an unknown parameter of the Hamiltonian. 
More specifically, we apply the soft actor-critic (SAC) RL algorithm~\cite{haarnoja_2019a} to the optimal control of a spin magnetometer. 
Once trained, we evaluate the performance of the agent in situations in which the Hamiltonian parameters differ from those seen in training.
An agent that generalizes well can be applied across a variety of scenarios without the cost of retraining, drastically reducing computational resources. 
This is beneficial for quantum sensors deployed in the field, where environmental factors change in time, along with the parameter to be estimated.

The paper is structured as follows.
We start in Sec.~\ref{s:system} by introducing the system being investigated, a spin magnetometer controlled by an external field through a series of parameterized pulses. 
In Sec.~\ref{s:rl}, we detail how the SAC RL algorithm is applied to the optimization of this system, and in Sec.~\ref{s:results} we evaluate the performance of the algorithm to situations not seen in training. 
We conclude in Sec.~\ref{s:conclusion} with a discussion of the results and future outlook.


\section{Spin System}
\label{s:system}

\begin{figure*}
    \includegraphics{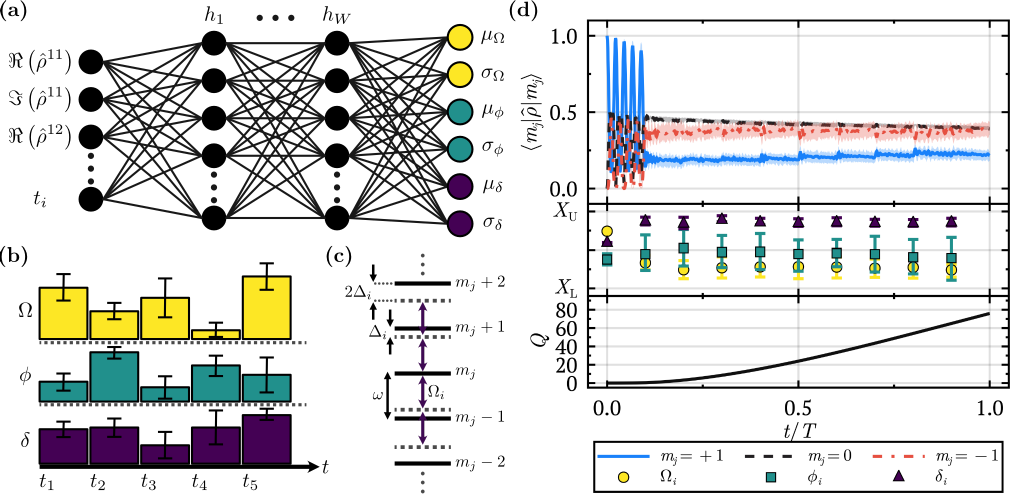}
    \caption{(a) Actor artificial neural network (ANN) architecture, see Appendix~\ref{s:app-sac}. The input layer takes the full density matrix $\hat{\rho}(t_i)$ at time $t_i$ prior to control pulse $i+1$, with real and imaginary components separated. The ANN has $W$ hidden layers, with the $n^{\mathrm{th}}$ layer denoted as $h_n$. The output layer neurons are interpreted as the mean $\mu$ and variance $\sigma$ of a normal probability distribution, from which the Hamiltonian control parameters $\left( \Omega, \phi, \delta \right)$ are sampled. (b) At time $t_i$, the reinforcement learning (RL) agent receives an observation from the environment, based on which a set of control parameters is sampled and applied as a pulse with fixed duration $\Tpulse$. After simulating the result, a sequence of such pulses is generated by repetition. (c) Energy level diagram for a spin-$j$ system, with magnetic sub-levels $m_j \in -j, \ldots, j$ uniformly separated by frequency $\omega$. An applied transverse oscillating field couples adjacent levels, with amplitude $\Omega$, phase $\phi$, and frequency $\delta$ detuned from resonance by $\Delta_i = \omega - \delta_i$. (d) Example RL agent policy for a $j=1$ system with $N=10$ control pulses. From a single trained agent, 100 trajectories are sampled. Average trajectories are shown, with colored bands or errorbars indicating the interquartile range (IQR). The top panel shows the average population in each spin sublevel as they evolve in time, while the middle panel shows the control parameters $X$ as they change with the state of the system, within their respective bounds $X \in \left[ X_{\mathrm{L}}, X_{\mathrm{U}} \right]$. The bottom panel shows the corresponding quantum Fisher information $Q$ as a function of time.}
    \label{fig:agent}
\end{figure*}

As a benchmarking model for optimizing the control of quantum systems with reinforcement learning (RL), we consider a spin with $d=2j+1$ levels, subjected to a background magnetic field whose magnitude we aim to measure with high sensitivity. 
The field is applied along the $z$-axis, with Larmor frequency $\omega$. 
The energy level scheme is depicted in Fig.~\ref{fig:agent}(c). 
The behavior of the spin systems is described by the Hamiltonian $\hat{H} = \omega \Sop{z}$, with $\Sop{q}$ the angular momentum operator along direction $q\in\left\{x, y, z\right\}$. 
In order to determine the magnitude $\omega$ of the magnetic field with high sensitivity, we apply a typical sensing procedure consisting of four steps. 
First, we prepare the initial state described by the density matrix $\rhoin$, which we assume to be independent of the parameter $\omega$. 
In the second step, the parametrization, the state evolves according to the Hamiltonian, encoding information about its parameters. 
This information is interrogated through measurement in a third step, after which the parameter may be estimated in step four. 
Improving the measurement sensitivity requires careful optimization of each step in this procedure, which is made difficult by limitations in available resources or control capabilities. 
Here, we focus on optimizing the state preparation and parametrization processes.

We start with considering the parametrization process, during which the initial state evolves so that the state $\hat{\rho}\left(t\right)$ after some time $t$ encodes information about the parameter of interest $\omega$. 
The amount of information about $\omega$ that is encoded in $\hat{\rho}\left(t\right)$ is quantified by the quantum Fisher information (QFI), 
\begin{equation}
    \label{eq:qfi}
    Q = \mathrm{tr} \left( \hat{\rho} \hat{L}^2 \right).
\end{equation}
Here, $\hat{L}$ is the symmetric logarithmic derivative (SLD) that is related to the derivative of the state with respect to the parameter of interest~\cite{liu_2020},
\begin{equation}
    \partial_{\omega} \hat{\rho} = \frac{1}{2} \left( \hat{\rho}\hat{L} + \hat{L}\hat{\rho} \right),
\end{equation}
with $\partial_x = \partial / \partial x$. 
Numerous efficient methods have been developed to compute the SLD, such as spectral methods, or through the Lyapunov or Saf\'{a}nek representations~\cite{liu_2020}.
In this work, we use the latter to numerically evaluate the SLD.
The theoretical limit of measurement sensitivity is given by the Cram\'er-Rao bound, which states that the maximum precision with which a parameter can be estimated is inversely proportional to the QFI~\cite{degen_2017, liu_2020}.
It is therefore the goal in quantum sensing to maximize the QFI of the state throughout its evolution. 
In contrast to the classical Fisher information, which depends on the measurements performed, the QFI assumes the choice of optimal measurements~\cite{liu_2020}. 
It thus quantifies an upper bound in information obtainable by an observer. 
In practice, the optimal measurement basis may be determined by the eigendecomposition of the SLD.

In the absence of decoherence, the optimal probe state, $\rhoopt$, is an equal superposition of the states with the maximal and minimal eigenvalues of $\partial_{\omega} \hat{H}$, \ie,\ of the magnetic sub levels $m_j = \pm j$. 
However, under the effects of decoherence or coherent noise processes, the system will deviate from this optimal state, and control fields are necessary to maintain optimality~\cite{fiderer_2019, jing_2015}. 
Additionally, in the case of multi parameter estimation, where the optimal probe state may differ for each parameter to be estimated, a compromise is required~\cite{goldberg_2021}. 
It is our goal to determine the optimal control fields to apply, as they vary in time, in order to maximize the QFI of the system.

Here, we parametrize the control in the form of a transverse field applied in the $x$-direction, with effective strength $\Omega$, that oscillates with frequency $\delta$ and phase $\phi$. 
The transverse field is applied in $N$ pulses, each of duration $\Tpulse$, and the parameters $\left(\Omega, \delta, \phi\right)$ are constant throughout each pulse. 
In the frame rotating with the oscillating field, the Hamiltonian for the $i^{\mathrm{th}}$ pulse is,
\begin{equation}
    \label{eq:ham}
    \hat{H}_i = \Omega_i \left[ \sin\left(\phi_i\right) \Sop{x} - \cos\left(\phi_i\right)\Sop{y} \right]
        + \left(\omega - \delta_i\right) \Sop{z}.
\end{equation}
The system evolves over a fixed duration $T=N\Tpulse$. 
This control parametrization generates transformations in $SU(2)$. 
For $j=1/2$, it is capable of producing arbitrary unitary transformations of the state (up to a global phase), so that the optimal probe state $\rhoopt$ can be prepared from any initial state. 
However, for spins $j > 1/2$, this is no longer the case and a Hamiltonian with a higher symmetry and more parameters, such as the generators of $SU(2j+1)$, would be necessary. 
As a result, for spins $j >1/2$, the applied control fields are insufficient for arbitrary control.
This is a realistic limitation in spin-based magnetometers, such as those based on nitrogen-vacancy centers in diamond, liquid nuclear magnetic resonance systems, or atomic ensembles. 
Thus, the optimal probe state $\rhoopt$ may not be obtainable and the achievable precision depends strongly on the choice of the initial state, making the optimization non-trivial.

The dynamics of open quantum systems are governed by the Gorini–Kossakowski–Sudarshan–Lindblad (GKSL) master equation,
\begin{equation}
    \label{eq:gksl}
    \partial_t \hat{\rho} = -\frac{i}{\hbar} \left[ \hat{H}, \hat{\rho} \right]
    + \sum_i \gamma_i \left[
        \hat{\Gamma}_i \hat{\rho} \hat{\Gamma}_i^{\dagger}
        - \frac{1}{2} \left\{ \hat{\Gamma}_i^{\dagger} \hat{\Gamma}_i,\hat{\rho} \right\}
    \right].
\end{equation}
The decoherence is expressed through the decay channels $\hat{\Gamma}_i$, which are jump operators describing the specific system's interaction with the environment; $\gamma_i$ are the corresponding decay rates. 
We consider a dephasing channel of the form $\hat{\Gamma}_1 = \Sop{z}$, and lossy excited states of the form $\hat{\Gamma}_2 = \Sop{-}$, with $\Sop{-}=\Sop{x}-i\Sop{y}$ the conventional angular momentum ladder operator. 
The decay rates associated with each decoherence channel are chosen to be equal, $\gamma_1 = \gamma_2 = \gamma$.
The density matrix $\hat{\rho}(t)$ evolves in time according to the chosen set of control pulses, $\left\{ \left(\Omega_i, \phi_i, \delta_i\right) \vert\ i = 1, 2, \ldots, N \right\}$ that define the Hamiltonian $\hat{H}$. 

We introduce the additional complication that the initial state of the system is not the optimal probe state $\rhoopt$, but rather an experimentally accessible spin-polarized state,
\begin{equation}
    \label{eq:ini-state}
    \rhoin = \ket{m_j = j}\bra{m_j=j}.
\end{equation}
Optimizing the control applied to such an initial state encompasses both the state-preparation and parametrization stages of the typical quantum sensing pipeline. 
Following these stages, the resulting QFI of the state determines the ultimate sensitivity with respect to the background field amplitude, $\omega$. 
The measurement protocol for obtaining this limit is itself an optimization problem, informed by the eigendecomposition of the SLD. 
Here, we only concern ourselves with finding a set of control parameters that maximizes the QFI in Eq.~(\ref{eq:qfi}).


\section{Reinforcement Learning}
\label{s:rl}

Rather than performing a conventional optimization of control parameters, we approach this problem with reinforcement learning (RL), which has already been used with success on related tasks in quantum sensing~\cite{huang_2024, schuff_2020, cimini_2023, chih_2024}. 

In contrast to conventional optimization, RL approaches replace the parameterized objective function with sequences of actions, taken by an agent, and rewards, provided by an environment in which the agent acts. 
Actions $a$ taken by the agent are sampled from its policy, $\pi$, which maps observations of the environment's state, $s$, to a probability distribution over the action space, $\pi:s\rightarrow p\left(s\right)$. 
Actions are then randomly sampled from the policy, $a \sim \pi(s)$. Depending on the state $s$ of the environment, the chosen action $a$, and the new state $s^{\prime}$ of the environment following the action $a$, a reward $R(s, a, s^{\prime})$ is provided. 
We then aim to determine the optimal policy $\piopt$ that maximizes the expected cumulative reward over a sequence of actions in an environment. 
Algorithms in RL typically approximate the policy with an artificial neural network (ANN) that takes observations from the environment as input and outputs parameters that define the policy as a parametrized probability distribution, as depicted in Fig.~\ref{fig:agent}(a).
The agent learns iteratively over episodes in which it interacts with the environment.
Received rewards are then back-propagated to update the network parameters such that the cumulative reward over the entire sequence of actions is maximized. 
Of the numerous RL algorithms suited to this task, we focus on the so-called soft actor-critic (SAC) method, an off-policy method defined for continuous or discrete action spaces~\cite{haarnoja_2019a}.
The SAC method boasts a relative insensitivity to hyperparameters, with stable convergence over many episodes. 
A detailed description of this algorithm is included in Appendix~\ref{s:app-sac}.

In the example of precisely estimating the magnitude of a magnetic field acting on a spin system as described in Sec.~\ref{s:system}, the spin system itself provides the RL environment. 
The agent observes the environment by considering the elements of the density matrix $\hat{\rho}\left(t_i\right)$ describing the spin system at time $t_i$.
As depicted in Fig.~\ref{fig:agent}(a), the real and imaginary parts of $\hat{\rho}$ are separated for more stable performance.
The output policy is a probability distribution over each external control field parameter, $\left( \Omega_i, \phi_i, \delta_i \right)$, at time $t_i$, as depicted in Fig.~\ref{fig:agent}(b). 
All control field parameters are continuous and may be bounded to intervals according to experimental or physical constraints, or reasonable ranges wherein the optimal value should exist.
The output layer of the so-called actor ANN (see Appendix~\ref{s:app-sac}) returns a mean and standard deviation for a normal distribution for each parameter $\Omega$, $\phi$, and $\delta$. 
To constrain these parameters to their respective boundaries, the normal distributions are clamped by $\tanh$, so that each component of the sampled action $a = \left(\tilde{\Omega}_i, \tilde{\phi}_i, \tilde{\delta}_i \right)$ is bounded by $\left(-1, 1 \right)$. 
Here, $\tilde{\Omega}$, $\tilde{\phi}$, and $\tilde{\delta}$ represent the parameters sampled from the output distribution of the actor ANN. 
The field parameter $X_i\in\left\{\Omega_i,\ \phi_i,\ \delta_i\right\}$ is scaled to the appropriate bounds by,
\begin{equation}
    \label{eq:param-scale}
    X_i = \frac{\left( X_{\mathrm{U}} - X_{\mathrm{L}} \right) \left( \tilde{X}_i + 1 \right)}{2} + X_{\mathrm{L}},
\end{equation}
with $X_{\mathrm{L}}$, $X_{\mathrm{U}}$ the lower and upper bound, respectively. 
The evolution of the state is simulated using the GKSL master equation, Eq.~(\ref{eq:gksl}), according to the corresponding control fields, returning a new state $\hat{\rho} \left( t_i + \Tpulse \right)$ after applying the pulse of duration $\Tpulse$. 
For the reward provided to the agent after each action, the change in the QFI after applying the pulse, $\Delta Q_i=Q_{i}-Q_{i-1}$, is a natural choice. 
The cumulative reward after $N$ steps is then the final QFI of the system, $Q=Q(T)$. 
Training over many episodes, the agent determines the optimal policy so as to maximize the QFI in time. 
A complete description of the SAC algorithm and the chosen network architecture is included in Appendix~\ref{s:app-sac}.


\section{Results}
\label{s:results}

\begin{figure}
    \includegraphics{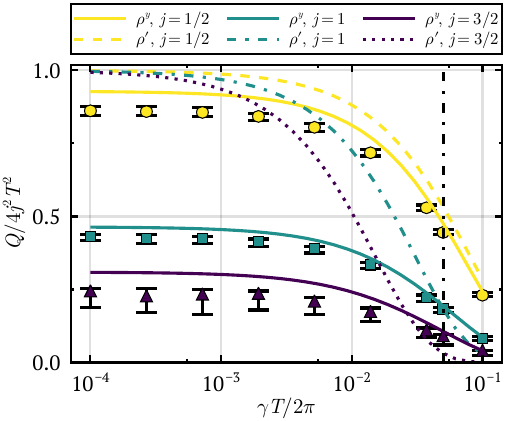}
    \caption{Performance of RL agents for different decoherence rates $\gamma$ when trained with a fixed decoherence rate of $\gamma=0.05 \times 2\pi/T$ (indicated by the dashed vertical line), for different spin manifolds $j$ (different colors). For each data point, 100 trajectories are sampled from the agent, and the median final quantum Fisher information (QFI) is plotted with errorbars indicating the interquartile range. Since the maximum theoretical QFI, without decoherence, scales with $4j^2 T^2$~\cite{jing_2015}, we have normalized the QFI by this factor to show the relative performance as $j$ is varied; a value of 1 on the above plot is therefore the highest possible QFI for this system. In each case, results are compared to the optimal probe state $\rhoopt$, and a strategy in which a single $\pi/2$-pulse is applied, preparing the system in a spin eigenstate (denoted $\rhoy$). Parameters of the RL agent networks were not updated during these evaluations, hence it was not permitted to adapt to the new data.}
    \label{fig:spin-decoherence}
\end{figure}

To investigate the performance of the introduced RL algorithm for optimizing the control parameters of a spin system in an external field by maximizing the QFI, we compare the QFI achieved after a sequence of $N$ pulses to the QFI of the optimal probe state $\rhoopt$, as discussed in Sec.~\ref{s:system}.
In the absence of decoherence, this probe state sets the ultimate upper bound on the QFI of the system.
In the presence of decoherence, this is no longer the case, since strategies can be developed to reduce the effects of decoherence, which produces superior results, as shown in Fig.~\ref{fig:spin-decoherence}.
However, the QFI of the optimal probe state still establishes a reasonable goal for the agent to reach and provides a method to determine the quality of the sampled control sequence.

Additionally, as discussed in Sec.~\ref{s:system}, the control Hamiltonian in Eq.~(\ref{eq:ham}) generates transformations in $SU\left(2\right)$ and provides only limited control capabilities for spins with $j>1/2$.
In those cases, given the chosen initial state $\rhoin$ in Eq.~(\ref{eq:ini-state}), the control is not necessarily sufficient to produce the optimal state $\rhoopt$.
In order to account for these disadvantages, we further compare the agent's performance against a chosen strategy that we consider to be optimal~\cite{jing_2015}. 
From the same initial state, $\rhoin$, we apply a single resonant $\pi/2$-pulse with $\delta=\omega$, $\phi=0$, and $\Omega=\pi / 2 \Tpulse$, preparing the system in an eigenstate of $\Sop{y}$, which we denote as $\rhoy$. 
In order for this strategy to be useful, the precise value of $\omega$ needs to be known to set the resonance condition of the pulse. 
This strategy sets a reasonable optimum for the agent to reach within its control capabilities, regardless of the spin manifold $j$. 
In the case of $j=1/2$, even though $\rhoy = \rhoopt$, when starting in the sub-optimal state $\rhoin$, the obtained total QFI $Q$ is lowered due to the time spent in sub-optimal states during the initial $\pi/2$-pulse.

During training, the environment parameters are chosen around the realistic assumption that the background field $\omega$ is much larger than the typical Rabi frequency of the control field.
We denote a reference Rabi frequency as $\OmegaRef = 2\pi$, from which all other system parameters are scaled.
Unless otherwise stated, the control amplitude is set within the bounds $\Omega \in \left[0, 5 \OmegaRef \right]$, with $\omega / \OmegaRef = 1000 \sqrt{2}$.
The phase of the control field is restricted to $\phi \in \left[0, \pi/2 \right]$, and its frequency to $\delta \in \left[ 0.99 \omega, 1.01 \omega \right]$.
The total duration of each sequence is $T = 10 \times 2\pi / \OmegaRef$, with $N=10$ pulses, and therefore a pulse duration of $\Tpulse = T/N$. The decoherence rate is $\gamma = 0.05 \times 2\pi/T$.

A major advantage of the introduced RL algorithm for parameter optimization compared to conventional optimization algorithms such as GRAPE~\cite{khaneja_2005} is the potential to generalize to scenarios not seen during the training process. 
Conventional optimization algorithms produce strict strategies that are only applicable to the specific choices of fixed system parameters for which they are optimized.
In contrast, an RL agent can potentially adapt to scenarios not seen in training and can thus be re-used in other situations.
This provides the flexibility to train the RL agent on an educated guess of the system parameters such as the decoherence rate, whose exact values in experimental realizations are likely to vary from theoretical predictions.

To investigate the ability of the RL agent to generalize to unseen scenarios, we vary the decoherence rate $\gamma$ and the spin quantum number $j$. 
We train one agent for each choice of $j\in\left\{1/2, 1, 3/2\right\}$ while using the same ANN architectures for all cases except for the input layer whose size depends on the density matrix, see Sect.~\ref{s:rl}.
Each ANN is randomly initialized, with network parameters for each layer sampled from a uniform distribution $\mathcal{U}\left( -1/\sqrt{k}, 1/\sqrt{k} \right)$, where $k$ is the number of input features to the network layer.
For training, we fix the decoherence rates to $\gamma = 0.05 \times 2\pi/T$.
We then evaluate each agent's policy with different values of $\gamma$, sampling $100$ trajectories for each. 
In Fig.~\ref{fig:spin-decoherence} we plot the median QFI obtained from these trajectories, along with the interquartile range (IQR) of the results.
We compare the results of the RL agent with the optimal probe state $\rhoopt$ and the state $\rhoy$ obtained with the strategy of applying a single resonant $\pi/2$-pulse.
For better comparison, we scale all QFI values by a factor of $4j^2T^2$ that describes the scaling behavior of the maximum theoretical QFI without decoherence~\cite{jing_2015}.
The performance of the RL agent follows the behavior of the $\pi/2$-pulse strategy even for decoherence rates that were not observed during the training process.
For higher values of $\gamma$, the strategies chosen by the RL agent surpass the optimal probe state, $\rhoopt$, which is due to the outermost spin levels being affected most strongly by the decoherence terms, see Eq.~(\ref{eq:gksl}).

\begin{figure}
    \includegraphics{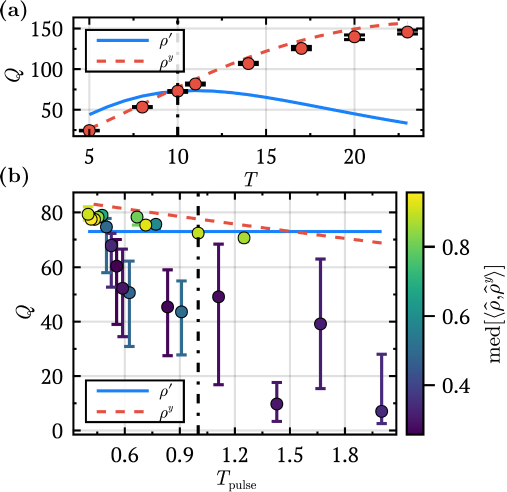}
    \caption{Performance of a reinforcement learning agent evaluated for a varied number of control pulses after training with $N=10$ pulses (indicated by the dashed vertical line) of fixed duration $T_{\mathrm{pulse}}$ and total duration $T$ for $j=1$. Each data point shows the median of $50$ sampled trajectories from the agent, with error bars indicating the interquartile range, and is compared to  results using the optimal probe state, $\rhoopt$, and the single $\pi/2$-pulse strategy, $\rhoy$. (a) The number of pulses is changed, while keeping $T_{\mathrm{pulse}}$ fixed; hence, the quantum Fisher information varies with the total simulation duration $T=NT_{\mathrm{pulse}}$. (b) The number of pulses $N$ is changed while keeping the total duration $T$ fixed. The pulse duration in each case is set as $T_{\mathrm{pulse}} = T/N$. Samples from the agent are colored according to the median inner product between the state after the first sampled pulse (with $\phi=0$), $\hat{\rho}$, and the $+1$ eigenstate of $\Sop{y}$, $\hat{\rho}^y$. How close the system is to such a state after this first pulse is strongly indicative to the overall performance of the agent, See Sec.~\ref{s:results}.}
    \label{fig:pulses}
\end{figure}

Next, we investigate the performance of the agent as it generalizes to a varied total duration $T$.
We train an agent with spin quantum number $j=1$ on $N=10$ pulses using a pulse duration of $\Tpulse$, and a total duration of $T = N \Tpulse$.
The performance of the agent is then evaluated for different times $T$ by varying the number of applied pulses $N$, each with the same pulse duration $\Tpulse$ used in training.
In Fig.~\ref{fig:pulses}(a), we plot the final QFI obtained using different durations $T$ and compare it with the optimal probe state $\rhoopt$ and the state $\rhoy$ obtained with the single $\pi/2$-pulse strategy.
The results follow those of the $\pi/2$-pulse strategy closely, demonstrating the ability of the agent to generalize.

While so far the pulse duration $\Tpulse$ has been fixed, we next vary the number of pulses while keeping the total duration $T=N\Tpulse$ fixed.
The pulse durations therefore change as $\Tpulse = T/N$.
The results are shown in Fig.~\ref{fig:pulses}(b), where we again compare the QFI achieved by the RL agent to that of the optimal probe state $\rhoopt$, and the state $\rhoy$ following a single $\pi/2$-pulse.
The agent is trained with a particular pulse duration, $\Tpulse$, and we find that it does not generalize well to changes in this parameter.
For most pulse durations $\Tpulse$, the achieved QFI is well below those found in the comparison states.
However, for certain choices of $\Tpulse$, the RL agent generalizes well and obtains QFIs that surpass the optimal probe state.

The varying performance of the RL agent in generalizing across different pulse durations can be explained by the fact that the chosen network architecture takes the time of the pulse $t_i$, but not the pulse duration as input, as discussed in Sec.~\ref{s:rl}. 
In its first action, the agent's policy is therefore incorrectly assuming that the pulse duration is the same as the one seen in the training process.
However, with a different duration, the state reached at the end of this initial pulse can be far from the anticipated optimal state.
While the agent can recover in the following decisions, this initial pulse is the most important for determining the final QFI. 
To demonstrate the strong effect of this first pulse, we sample $100$ initial actions from the agent, and simulate the state of the system $\hat{\rho}$ after the initial pulse with duration $\Tpulse = T/N$, fixing the phase to $\phi_1=0$.
The points in Fig.~\ref{fig:pulses}(b) are colored by the median of the Frobenius inner product $\left\langle\hat{\rho},\rhoy\right\rangle$ of the state $\hat{\rho}$ after application of the first pulse with the state $\rhoy$ reached after a single $\pi/2$ pulse, indicating the amount of overlap between the two states, akin to fidelity.
The chosen pulse duration impacts the overlap between the two states $\hat{\rho}$ and $\rhoy$ and it can clearly be seen that the closer the state after the initial pulse is to $\rhoy$, the higher the QFI reached by the agent.
Thus, the learned policy shows a strong dependence on the pulse duration and the agent can only generalize well to situations where the first applied pulse leads to a similar state as experienced in the training process.

\begin{figure*}
    \includegraphics{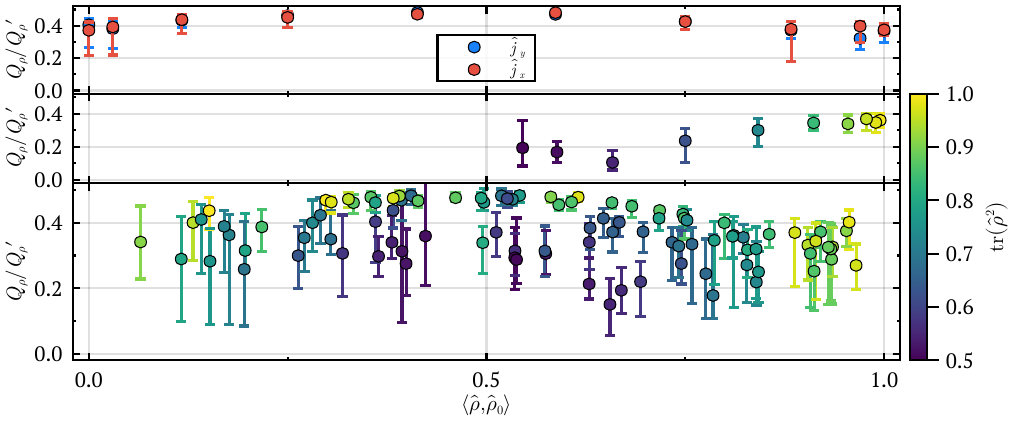}
    \caption{Performance of the RL agent as the initial state of the system, $\hat{\rho}$, is varied from that used in training, $\rhoin$, for $j=1$. For each initial state, the median quantum Fisher information (QFI) across 50 sampled trajectories is shown, with the interquartile range indicated by the errorbars. In all cases, the final QFI $Q$ is normalized by the maximum possible for that state, $Q_{\rho}^{\prime}$~\cite{fiderer_2019}. (Top) $\hat{\rho}$ is transformed by $\hat{U} = \exp \left( -i \theta \Sop{q} \right)$ with $q=x, y$, and $\theta \in [0, \pi]$, demonstrating how the agent responds to initial states that are spin polarized along different axes than previously seen. (Middle) initial states are reduced in purity, $\mathrm{tr}\left( \hat{\rho}^2 \right)$, by adding a mixing term $\propto \identity$ with varied amplitude ($\mathrm{tr}\left( \hat{\rho} \right)$ is renormalized in each case). Note that as the purity changes, so too does $\langle \hat{\rho}, \rhoin \rangle$. (Bottom) 100 states are randomly generated through a combination of the previous two approaches. The initial state is transformed through $\hat{U} = \exp \left[ -i \left( \theta_x \Sop{x} + \theta_y \Sop{y} \right) \right]$ with $\theta_x$, $\theta_y$ chosen randomly, and with a random amount of mixing added afterwards.}
    \label{fig:states}
\end{figure*}

We next investigate the ability of the agent to generalize to different initial states $\rhoin$ from which the generated trajectories start.
During the training process, the agent always receives the same initial state $\rhoin=\left|m_j=j\right\rangle\left\langle m_j=j\right|$.
However, it experiences a variety of states throughout each episode depending on the chosen actions to manipulate the given state.
We now evaluate the trained agent's performance when given different initial states.
First, we explore different spin-polarized states, which we generate by rotating the initial state $\rhoin$ coherently around either the $x$- or $y$-axis. 
More specifically, the initial state is transformed according to $\hat{\rho} = \hat{U}^{\dagger} \rhoin \hat{U}$ with $\hat{U} = \exp \left( -i \theta \Sop{q} \right)$, $\theta \in [0, \pi]$, and $q \in \left\{ x, y\right\}$. 
Fig.~\ref{fig:states} shows the resulting QFI for $j=1$, where each initial state is scaled by the corresponding theoretical upper bound $Q_{\rho}^{\prime}$, assuming no decoherence~\cite{fiderer_2019}. 
We characterize the initial states through the Frobenius inner product $\left\langle\hat{\rho},\rhoin\right\rangle$ with $\rhoin$, Eq.~(\ref{eq:ini-state}), indicating the distance from the initial state used during the training process. 
The results show that the agent generalizes well and performs notably better around $\left\langle\hat{\rho},\rhoin\right\rangle=0.5$, which is the case when the initial state $\hat{\rho}$ is close to eigenstates of $\Sop{x}$ or $\Sop{y}$.
Thus, the agent generalizes best for initial states close to the training initial state $\rhoin$, but shows a strong performance throughout all tested states.
The maximum QFI the agent reaches is found to be about half of the upper bound $Q_{\rho}^{\prime}$ of the corresponding state.
This upper bound is determined without the presence of decoherence effects, while those effects are taken into account in the agent's trajectory.
The agent is thus not expected to reach this upper bound, as decoherence effects reduce the QFI over time.
A ratio of $Q_{\rho}/Q_{\rho}^{\prime}\approx 0.5$ is the highest value reached in our simulations, based on the chosen Hamiltonian parameters, and is thus considered as a successful generalization.

Next we explore the generalizability when varying the purity $\mathrm{tr}\left(\hat{\rho}^2\right)$ of the initial state.
For this, we add the identity matrix $\identity$ to $\rhoin$ and scale its influence by a factor while keeping the trace of the resulting state normalized.
The results are presented in the middle panel of Fig.~\ref{fig:states}, where the QFI is again normalized by the theoretical maximum $Q^{\prime}_{\rho}$ for each state $\hat{\rho}$.
The change in purity also reflects in a change in the Frobenius inner product with the training initial state $\rhoin$.
The color scale in the lower two panels of Fig.~\ref{fig:states} shows the purity of each initial state, clearly demonstrating the agent's limitation when generalizing to more strongly mixed states.

In the lower panel of Fig.~\ref{fig:states}, we combine the two previous approaches and generate $100$ random spin-polarized states with random purity.
For each randomly chosen setting, the initial state $\rhoin$ is transformed by $\hat{U}=\exp\left[-i\left(\theta_x\Sop{x}+\theta_y\Sop{y}\right)\right]$, with $\theta_x,\ \theta_y\in\left[0,2\pi\right)$.
The purity is then varied by adding the identity matrix with a randomly chosen scaling factor.
The randomly chosen initial state is again identified by the Frobenius inner product $\left\langle\hat{\rho},\rhoin\right\rangle$ with $\rhoin$ and the purity of each state is indicated by the color scale.
While a relative maximum in performance is observed around $\left\langle \hat{\rho},\rhoin\right\rangle\approx0.5$, it shows a fairly consistent behavior over varied spin polarization.
At the same time, the performance is negatively influenced when the purity of the state falls below $\mathrm{tr}\left(\hat{\rho}^2\right)\lesssim 0.75$.
This observed behavior is a clear combination of the two individual effects found in the upper two panels of Fig.~\ref{fig:states} and verifies that, while generalizing well for different polarizations of the initial state, the agent's strategy does not perform well on strongly mixed states.


\section{Conclusion}
\label{s:conclusion}
Our results demonstrate that reinforcement learning (RL) is a promising approach for optimizing the control of quantum systems.
Considering the specific example of a spin subjected to a magnetic field, we successfully train an RL agent to perform actions in the form of applying pulses of transverse fields with varying strength, oscillation frequency, and phase, such that the sensitivity of sensing the strength of the magnetic field is optimized.
We further show that, once trained, the agent is able to generalize to situations that were not considered during the training process.
Such situations include a varying decoherence rate (Fig.~\ref{fig:spin-decoherence}), different numbers of pulses and different pulse durations (Fig.~\ref{fig:pulses}), or changes to the initial state of the system (Fig.~\ref{fig:states}).

While promising generalizability was observed when varying the decoherence rate or the number of pulses, changes to the pulse duration resulted in mixed performance.
This effect is mainly caused by the agent expecting its choice of control amplitude $\Omega$ to result in the same pulse area seen in the training process.
For varying pulse durations, the state following the first action taken by the agent can deviate strongly from the expected state based on the training process, and is predictive of the final outcome.
To achieve an increased performance for these cases, the pulse duration can be added as input to the ANN used in the RL algorithm, which is planned for future works.

When considering changes to the initial state of the system either by rotating it around the $x$- or $y$-axis or by decreasing its purity, the agent generalizes best to states above a particular purity threshold, and when the initial state is already close to a magnetically sensitive state.
Improvement in the generalizability is similarly expected by providing additional information to the ANN during training and evaluation.
This might be achieved by either providing more information as additional inputs to the ANN, or by providing more diverse scenarios during the training process.

While there remains great potential for exploring various configurations of system parameters, our initial results indicate that RL is well suited to complicated quantum optimal control problems.
We specifically use the soft actor-critic (SAC) method to implement the RL algorithm (see Appendix~\ref{s:app-sac}), whose convergence to near-optimal solutions is stable over multiple runs and insensitive to the choice of hyperparameters in the invovled ANNs.
Furthermore, the SAC algorithm is agnostic to the details of the physical system, in that the same network architectures and their corresponding hyperparameters can be used with similar effect in different systems, with their own action spaces and control parametrizations.
This robustness and the proven potential for generalization to different system parameters make the SAC-based RL algorithm outperform conventional optimizers which commonly produce a strict strategy only applicable to the specific parameters it was optimized for.
While the training of the RL algorithm is generally more computationally expensive than most conventional optimizers, the chosen ANNs in this work are small enough to allow the usage of consumer-grade hardware.
The primary computational bottleneck are the simulations of the physical system during the training process.
Once the agent is trained, the evaluation of the policy is fast and inexpensive.

Overall, our study demonstrates various strengths of SAC-based RL algorithms when solving a quantum optimal control problem in situations beyond those experienced during the training stage.
The agents are able to perform well when optimizing the spin-based magnetometer for a realistic set of parameters, including the presence of decoherence and limited control capabilities.
Based on the results in this toy-model system, we anticipate that such algorithms perform well in more complicated environments, where the training process may need to be more diverse, or include additional observations as inputs to the ANNs.
It is of additional interest to investigate the agent's ability to adapt to dynamic environments in real-time, which would allow it to learn as it acquires new experiences.
This approach requires further exploration and is left for future works.
Similarly, there is great potential for exploring situations where quantum resources, such as entanglement, play a crucial role in optimizing performance, as in quantum sensing applications, or situations where multiple parameters are being estimated simultaneously.


\section*{Acknowledgments}
The authors thank Benjamin MacLellan for valuable discussions.
The authors acknowledge the support of the Natural Sciences and Engineering Research Council of Canada (NSERC) [funding reference number ALLRP 578468 - 22].
This research was enabled in part by support provided by the Digital Research Alliance of Canada (alliancecan.ca).

\section*{Code Availability}
The code used to produce the presented results is available at \url{https://github.com/APRIQuOt/rl-spin-magnetometer}.
Simulations were performed in \texttt{Julia}~\cite{bezanson_2017}, through the \texttt{OrdinaryDiffEq.jl} package~\cite{rackauckas_2017}.
The RL implementation was done in \texttt{Python} using \texttt{PyTorch}~\cite{ansel_2024a}.
The \texttt{Julia} simulations were accessible from \texttt{Python} through the \texttt{JuliaCall}~\cite{PythonCall.jl} package.
Plots were generated in \texttt{Julia}.

\clearpage

\appendix
\renewcommand{\thefigure}{A\arabic{figure}}
\setcounter{figure}{0}
\label{s:supplemental}


\section{Soft Actor-Critic}
\label{s:app-sac}

The SAC method is an off-policy algorithm, known for its robust convergence, and insensitivity to hyperparameters~\cite{haarnoja_2019a}. One of the unique features of SAC is that it is entropy-regularized, where the reward is modified by the entropy of the policy. The entropy for the random variable $x$ sampled from a probability distribution $P$ is the expectation value of its log-probability,
\begin{equation}
    \label{eq:entropy}
    S(x) = \underset{x \sim P}{\mathbb{E}} \left[ - \log P(x) \right].
\end{equation}
For a policy $\pi$, the value of a state $s$ of the environment in SAC is,
\begin{multline}
    \label{eq:value-func}
    V^{\pi} (s) = \underset{\tau \sim \pi}{\mathbb{E}} \left\{
        \sum_{t=0}^{\infty} \gamma^t \left( \alpha R(s_t, a_t, s_{t+1}) \right. \right. \\
        \left. \left. + S\left[ \pi(\cdot | s_t ) \right] \right) \Bigg\vert s_0=s \right\},
\end{multline}
which is averaged over trajectories $\tau$ sampled from the policy. Each trajectory is a sequence of state-action pairs, $(s_t, a_t)$, for each step $t$ in the episode, where the action at each step is sampled from the policy, $a_t \sim \pi$. The value function is thus a balance between the reward $R$ and the entropy of the policy $S$. The reward $R$ is multiplied by the reward scaling $\alpha$, which is a hyperparameter that adjusts the relative emphasis between exploitation versus exploration. The parameter $\gamma^t$ is the reward discount factor, common in various other RL algorithms.

Based on the value function, Eq.~(\ref{eq:value-func}), the optimal policy $\piopt$ must maximize both the cumulative reward and the randomness of the policy, simultaneously. The optimal policy is,
\begin{equation}
    \label{eq:opt-policy}
    \piopt = \underset{\pi}{\mathrm{argmax}} \underset{\tau \sim \pi}{\mathbb{E}}
    \left\{
        \sum_{t=0}^{\infty} \gamma^t \left[ \alpha R(s_t, a_t, s^{\prime}_t) + S(\pi(\cdot \vert s_t))\right]
    \right\}.
\end{equation}
Modifying the reward with the entropy of the policy encourages both exploitation of the reward, and exploration of different actions, as weighted by the reward scaling $\alpha$. This balance tends to help the agent long-term, as the emphasis on randomness can help explore new actions throughout training, and prevent the model from getting stuck in local minima.

In order to learn and approximate the optimal policy $\piopt$ in Eq.~(\ref{eq:opt-policy}), the SAC method utilizes five ANNs, all related through the definition of their loss functions. The first ANN is the actor, which approximates the policy. More specifically, it takes in an observation $s$ from the environment, and returns the parameters of a probability distribution over the action space, such as the mean and standard deviation of a Gaussian distribution from which actions $a$ may be sampled. Another ANN is used to approximate the value function, Eq.~(\ref{eq:value-func}), called the value network, which takes in observations $s$ and predicts the expected reward. There is also a target value network, which is a soft copy of the value network with its own update rule, the role of which is discussed below.

As SAC is an actor-critic method, we require an ANN to represent the critic, which evaluates the performance of the actor given the state $s$ and action taken by the actor $a$. This ANN approximates the Q-function, common in many other Q-learning algorithms. The Q-function has a similar form as the value function, Eq.~(\ref{eq:value-func}), but does not include the entropy bonus from the initial state,
\begin{multline}
    \label{eq:q-func}
    Q^{\pi} (s, a) = \underset{\tau\sim\pi}{\mathbb{E}} \left\{
        \alpha \sum_{t=0}^{\infty} \gamma^t R(s_t, a_t, s_{t+1}) \right. \\
        \left. + \sum_{t=1}^{\infty} \gamma^t S\left[ \pi(\cdot | s_t) \right] \Bigg\vert s_0=s, a_0=a
        \right\}. 
\end{multline}
We make use of two critics, each an ANN approximating the Q-function above [Eq.~(\ref{eq:q-func})], in order to implement the clipped double-Q trick~\cite{hasselt_2010}. When computing the loss during training, the lowest estimated reward from the two critics is used, so that the performance of the actor is not over-estimated. This method is used in various different Q-learning approaches, showing increased stability in training. Together, the actor, two critics, value network, and target value network, provide everything needed to determine the optimal policy for a given environment and action space.

Training each of the networks mentioned above to correctly approximate their target functions is done by allowing the agent to interact with the environment, gaining new experiences. The SAC method makes use of a replay buffer $\mathcal{D}$, allowing it to learn from previous experiences even while the actions taken in those transitions came from an old policy. The replay buffer contains the set of all previous transitions, $\mathcal{D} = \left\{\left( s, a, r, s^{\prime}, d \right)\right\}$, including the observation $s$, action taken $a$, reward $r$, new observation $s^{\prime}$, and the terminal flag $d$, which is a Boolean indicating if $s^{\prime}$ is terminal or not, signaling the end of the episode. The replay buffer is used during training to approximate expectation values by sampling previous transitions, and averaging over them.

After each action taken by the agent, the network parameters are updated according to their loss functions. We start with the value network, for which the loss function is,
\begin{multline}
    \label{eq:v-loss}
    L_{\mathrm{value}} = \underset{s\sim\mathcal{D},\ \tilde{a}\sim\pi}{\mathbb{E}} \Bigg\{ V(s) - \Bigg[ \underset{i=1, 2}{\mathrm{min}} Q_i (s, \tilde{a}) \\
    - \log \pi(\tilde{a} | s) \Bigg] \Bigg\},
\end{multline}
where we denote that the actions used are newly sampled from the current policy, $\tilde{a} \sim \pi$, not from the replay buffer. This ensures that we can learn from past experiences even if they were made with poor policies. Note that in the above, we employ the clipped double-Q trick, as indicated by taking the minimum output of each critic, indexed by $i=1$, $2$. The loss of the value network is therefore related to the performance of the critic networks.

Next, each critic network is updated independently. The critic loss is the mean-squared-error between the value predicted by the $i^{\mathrm{th}}$ critic, $Q_i(s, a)$, and the estimated future reward, which is just the sum of the sampled reward $r$ and the estimated value of the new state $s^{\prime}$ as predicted by the target value network,
\begin{multline}
    \label{eq:critic-loss}
    L^i_{\mathrm{critic}} = \underset{(s, a, r, s^{\prime}, d) \sim \mathcal{D}}{\mathbb{E}} \Big\{ \left[ Q_i(s, a) \right. \\
    \left. - \left[ r + \gamma (1-d) V_{\mathrm{target}}(s^{\prime}) \right] \right]^2 \Big\}.
\end{multline}
In this case, states and actions are all sampled from the buffer. Similar to Eq.~(\ref{eq:critic-loss}), the loss of the critic networks therefore depends on the performance of the target value network.

The actor network approximates the policy, which attempts to maximize the expected cumulative reward. Its loss function is therefore the negative of this, where we invoke the current prediction from the critics (using the clipped double-Q trick),
\begin{equation}
    \label{eq:actor-loss}
    L_{\mathrm{actor}} = \underset{s\sim \mathcal{D},\ \tilde{a}\sim\pi}{\mathbb{E}} \left[ \log \pi(\tilde{a} | s) - \underset{i=1,2}{\mathrm{min}}Q_i (s, \tilde{a}) \right].
\end{equation}
Actions are again sampled from the current policy, rather than from the replay buffer. When back-propagating this loss through the actor network, since the sampled actions are stochastic, we use the re-parametrization trick to restore a deterministic dependence of $\tilde{a}$ on the actor network parameters~\cite{haarnoja_2019a}.

The target value network is the last to be updated. Its parameters, $\hat{\psi}_{\mathrm{target}}$, are a soft-copy of those of the value network's, $\hat{\psi}_{\mathrm{value}}$,
\begin{equation}
    \label{eq:target-loss}
     \hat{\psi}_{\mathrm{target}} \leftarrow \beta \hat{\psi}_{\mathrm{value}} + (1 - \beta) \hat{\psi}_{\mathrm{target}}.
\end{equation}
The hyperparameter $\beta$ is typically small, $\beta \approx 0.005$. The target value network is therefore a time-delayed soft-copy of the value network. It is used in computing the loss for the critic networks, Eq.~(\ref{eq:critic-loss}), but since it is the last to be updated, any changes to its parameters will not take effect until the next training step. This time delay is important in the stability of the training, and ties the performance of the other networks together. As mentioned previously, each of the agent ANNs is updated in the manner described above after each transition in an episode. After many such episodes, the SAC agent therefore converges such that its ANNs are good approximations of their corresponding functions.

Unless otherwise stated, all ANNs implemented here (Sec.~\ref{s:results}) are typical feed-forward architectures, with three inner layers of 256 neurons, and ReLU activation functions. We use the Adam optimizer for each, with learning rates $\varepsilon=0.001$, reward discounts $\gamma=0.99$, and target value network learning parameters $\beta = 0.005$. Generally, all of these choices for hyperparameters are subject to optimization, although the SAC algorithm boasts reasonable stability to changes in them. The important hyperparameter that the method is sensitive to is the reward scaling, $\alpha$, which controls the emphasis of randomness in the policy. If $\alpha$ is too small (encouraging randomness), the training can be unstable, but if it is too large (encouraging exploitation of the reward), the agent can get stuck in local optima and acquires less diverse experience to learn from. In this case, we found the best results for $2 \leq \alpha \leq 10$.

%

\end{document}